\documentclass[aps,pre,onecolumn,superscriptaddress,groupedaddress]{revtex4}

\usepackage[utf8]{inputenc}
\usepackage[english]{babel}
\usepackage{indentfirst}
\usepackage{fancyhdr}
\usepackage[final]{showkeys}
\usepackage{graphicx}
\usepackage{subfig}
\usepackage{amssymb}
\usepackage{latexsym}
\usepackage{amsthm}
\usepackage{amsmath}
\usepackage{eucal}
\usepackage{amsfonts}
\usepackage{mathtools}
\usepackage{pgf,tikz}
\usepackage{mathrsfs}
\usepackage{epstopdf}
\usetikzlibrary{arrows}
\usepackage{enumerate}
\usepackage{epigraph}
\usepackage[a4paper, left=1.5cm, right=1.5cm, top=2cm, bottom=2cm]{geometry}
\usepackage{floatpag}

\DeclarePairedDelimiter\bra{\langle}{\rvert}
\DeclarePairedDelimiter\ket{\lvert}{\rangle}
\DeclarePairedDelimiterX\braket[2]{\langle}{\rangle}{#1 \delimsize\vert #2}
\DeclarePairedDelimiterX\inner[2]{\langle}{\rangle}{#1,#2}
\DeclarePairedDelimiterX\ketbra[2]{\lvert}{\rvert}{#1 \delimsize\rangle\!\delimsize\langle #2}

\newcommand{\me}[1]{\left\langle #1 \right\rangle }

\begin{document}

\title{Quantum-heat fluctuation relations in $3$-level systems under projective measurements}

\author{G. Giachetti}
\email{ggiachet@sissa.it}
\affiliation{SISSA, Via Bonomea 265, I-34136 Trieste, Italy}
\affiliation{INFN, Sezione di Trieste, I-34151 Trieste, Italy}

\author{S. Gherardini}
\email{gherardini@lens.unifi.it}
\affiliation{SISSA, Via Bonomea 265, I-34136 Trieste, Italy}
\affiliation{Department of Physics and Astronomy \& LENS, University of Florence, via G. Sansone 1, I-50019 Sesto Fiorentino, Italy}

\author{A. Trombettoni}
\email{andreatr@sissa.it}
\affiliation{Department of Physics, University of Trieste, Strada Costiera 11, I-34151 Trieste, Italy}
\affiliation{CNR-IOM DEMOCRITOS Simulation Center and SISSA, Via Bonomea 265, I-34136 Trieste, Italy}

\author{S. Ruffo}
\email{ruffo@sissa.it}
\affiliation{SISSA, Via Bonomea 265, I-34136 Trieste, Italy}
\affiliation{INFN, Sezione di Trieste, I-34151 Trieste, Italy}
\affiliation{Istituto dei Sistemi Complessi, Consiglio Nazionale delle Ricerche,
via Madonna del Piano 10, I-50019 Sesto Fiorentino, Italy}

\begin{abstract}
We study the statistics of energy fluctuations in a three-level quantum system subject to a sequence of projective quantum measurements. We check that, as expected, the quantum Jarzynski equality holds provided that the initial state is thermal. The latter condition is trivially satisfied for two-level systems, while this is generally no longer true for $N$-level systems, with $N > 2$. Focusing on three-level systems, we discuss the occurrence of a unique energy scale factor $\beta_{\rm eff}$ that formally plays the role of an effective inverse temperature in the Jarzynski equality. To this aim, we introduce a suitable parametrization of the initial state in terms of a thermal and a non-thermal component. We determine the value of $\beta_{\rm eff}$ for a large number of measurements and study its dependence on the initial state. Our predictions could be checked experimentally in quantum optics.
\end{abstract}

\maketitle

\section{Introduction}

Fluctuation theorems relate fluctuations of thermodynamic quantities of a given system to equilibrium properties evaluated at the steady state\,\cite{Esposito2009,Campisi2011,SeifertRPP2012,DeffnerBook2019}. This statement finds fulfillment in the Jarzynski equality, whose validity has been extensively theoretically and experimentally discussed in the last two decades for both classical and quantum systems\,\cite{JarzynskiPRL1997,CrooksPRE1999,CollinNature2005,ToyabeNatPhys2010,Kafri2012,Albash13PRE88,Rastegin13JSTAT13,Sagawa2014,AnNatPhys2015,BatalhaoPRL2014,CerisolaNatComm2017,BartolottaPRX2018,Hernandez2019}.

The evaluation of the relevant work originated by using a coherent modulation of the system Hamiltonian has been the subject of intense investigation\,\cite{TalknerPRE2007,CampisiPRL2009,MazzolaPRL2013,AllahverdyanPRE2014,TalknerPRE2016,JaramilloPRE2017,DengENTROPY2017}. A special focus was devoted to the study of heat and entropy production, obeying of the second law of thermodynamics, the interaction with one or more external bodies, and/or the inclusion of an observer\,\cite{JarzynskiPRL2004,Campisi15NJP17,Campisi17NJP19,BatalhaoPRL,Gherardini_entropy,ManzanoPRX,Irreversibility_chapter,SantosnpjQI2019,KwonPRX2019,RodriguesPRL2019}.

In this respect, the energy variation and the emission/absorption of heat induced---and sometimes enhanced---by the application of a sequence of quantum measurements
was studied\,\cite{Campisi2010PRL,CampisiPRE2011,Yi2013,WatanabePRE2014,HekkingPRL2013,AlonsoPRL2016,GherardiniPRE2018,Hernandez2019}. In such a case, the term quantum heat has been used\,\cite{Elouard2016,GherardiniPRE2018}; we will employ it as well in the following to refer to the fact that the fluctuations of energy exchange are induced by quantum projective measurements performed during the time evolution of the system. As recently discussed in Ref.\,\cite{Hernandez2019}, the information about the fluctuations of energy exchanges between a quantum system and an external environment may be enclosed in an energy scaling parameter that only depends on the initial and the asymptotic (for long times) quantum states resulting from the system dynamics.

In Ref. \cite{GherardiniPRE2018}, the effect of stochastic fluctuations on the distribution of the energy exchanged by a quantum two-level system with an external environment under sequences of quantum measurements was characterized and the corresponding quantum-heat probability density function was derived. It has been shown that, when a stochastic protocol of measurements is applied, the quantum Jarzynski equality is obeyed. In this way, the quantum-heat transfer was characterized for two-level systems subject to projective measurements. Two-level systems have the property that a density matrix in the energy basis (as the one obtained after a measurement of the Hamiltonian operator\,\cite{TalknerPRE2007}) can be always written as a thermal state and, therefore, the Jarzynski equality has a $1$ on its right-hand side. Therefore, a natural issue to be investigated is the study of quantum-heat fluctuation relations for $N$-level systems, e.g., $N = 3$, where this property of the initial state of a two-points measurement scheme of being thermal is no longer valid. So, it would be desirable, particularly for the case of a large number of quantum measurements, to study the properties of the characteristic function of the quantum heat when initial states cannot be written as thermal states.

With the goal of characterizing the effects of having arbitrary initial conditions, in this paper, we study quantum systems described by finite dimensional Hilbert spaces, focusing on the case of three-level systems. We observe that finite-level quantum systems may present peculiar features with respect to continuum systems. As shown in Ref. \cite{JaramilloPRE2017}, even when the quantum Jarzynski equality holds and the average of the exponential of the work equals the free energy difference, the variance of the energy difference may diverge for continuum systems, an exception being provided by finite-level quantum systems.

In this paper we analyze, using numerical simulations, (i) the distribution of the quantum heat originated by a three-level system under a sequence of $M$ projective measurements in the limit of a large $M$, and (ii) the behavior of an energy parameter $\beta_{\rm eff}$, such that the Jarzynski equality has $1$ on its right-hand side, always in the limit of a large $M$. We also discuss the dependence of $\beta_{\rm eff}$ on the initial state, before the application of the sequence of measurements.

\section{The Protocol}

Let us consider a quantum system described by a finite dimensional Hilbert space. We denote with $H$ the time-independent Hamiltonian of the system that admits the spectral decomposition
\begin{equation}
H = \sum^N_{k=1} E_k \ketbra{E_k}{E_k} \ ,
\end{equation}
where $N$ is the dimension of the Hilbert space. We assume that no degeneration occurs in the eigenstates of $H$.

At time $t=0^{-}$, just before the first measurement of $H$ is performed,
the system is supposed to be in an arbitrary quantum state described by the density matrix $\rho_0$ s.t.\,$[H, \rho_0] = 0$. This allows us to write
\begin{equation} \label{inital}
\rho_{0} = \sum^N_{k=1} c_k \ketbra{E_k}{E_k} \ ,
\end{equation}
where $1 \geq c_k \geq 0 \ \forall k = 1, \dots, N$ and $\sum^N_k c_k =1$.

Then, we assume that the fluctuations of the energy variations, induced by a given transformation of the state of the system, are evaluated by means of the so-called two-point measurement (TPM) scheme\,\cite{TalknerPRE2007}. According to this scheme, a quantum projective measurement of the system hamiltonian is performed both at the initial and the final times of the transformation. This hypothesis justifies the initialization of the system in a mixed state, as given in Equation (\ref{inital}). By performing a first projective energy measurement, at time $t=0^{+}$, the system is in one of the states $\rho_n = \ketbra{E_n}{E_n}$ with probability $p_n = \bra{E_n} \rho_0 \ket{E_n}$, while the system energy is $E_n$. Afterwards, we suppose
that the system $S$ is subject to a number $M$ of consecutive projective measurements of the generic observable
\begin{equation}
O = \sum^N_{k=1} \Omega_{k} \ketbra{\Omega_k}{\Omega_k} \ ,
\end{equation}
where $\Omega_k$ and $\ket{\Omega_k}$ denote, respectively, the outcomes and the eigenstates of $O$. According to the postulates of quantum mechanics, the state of the system after one of these projective measurements is given by one of the projectors $\ketbra{\Omega_n}{\Omega_n}$. Between two consecutive measurements, the system evolves with the unitary dynamics generated by $H$, i.e., $U(\tau_i) = e^{-iH \tau_i}$, where $\hbar$ has been set to unity and the waiting time $\tau_i$ is the time difference between the $(i-1)^{\text{th}}$ and the $i^{\text{th}}$ measurement of $O$.

In general, the waiting times $\tau_i$ can be random variables, and the sequence $(\tau_1,\ldots,\tau_M)$ is distributed according to the joint probability density function $p(\tau_1,\ldots,\tau_M)$. The last (i.e, the $M^{\text{th}}$) measurement of $O$ is immediately followed by a second projective measurement of the energy, as prescribed by the TPM scheme. By denoting with $E_m$ the outcome resulting from the second energy measurement of the scheme, the final state of the system is $\rho_m = \ketbra{E_m}{E_m}$ and the quantum heat $Q$ exchanged during the transformation is thus given by
\begin{equation}
Q = E_m - E_n \ .
\end{equation}

As $Q$ is a random variable, one can define the characteristic function
\begin{equation}
G(\epsilon) \equiv \me{e^{-\epsilon Q}} =
\sum_{m,n}p_{m|n}p_{n}e^{-\epsilon(E_m - E_n)} \ ,
\end{equation}
where $p_{m|n}$ denotes the probability of obtaining $E_m$ at the end of the protocol conditioned to have measured $E_n$ at the first energy measurement of the TPM scheme. If the initial state is thermal, i.e., $\rho_0 = e^{- \beta H}/Z$, then one recovers the Jarzynski equality stating that
\begin{equation}
G(\beta) = \me{e^{-\beta Q}} = 1 \ .
\end{equation}

Let us notice that $G(\epsilon)$ is a convex function such that $G(0)=1$ and $G(\pm \infty) \rightarrow + \infty$, as discussed in\,Ref. \cite{Hernandez2019}. Hence, as long as $\frac{\partial G}{\partial \epsilon}(0) \neq 0$, one can unambiguously introduce the parameter $\beta_{\rm eff} \neq 0$ defined by the relation
\begin{equation}
G(\beta_{\rm eff}) = 1 \, ,
\end{equation}
which formally plays the role of an effective inverse temperature. Focusing on three-level systems, in the following, we will study the characteristic function $G(\epsilon)$ and the properties of such a parameter $\beta_{\rm eff}$. For comparison, we first pause in the next subsection to discuss what happens for two-level systems.

\subsection{Intermezzo on Two-Level Quantum Systems}

We pause here to remind the reader of the results for two-level systems. The state of any two-level system, diagonal on the Hamiltonian basis, is a thermal state for some value of $\beta$. Of course, if the state is thermal, the value of $\beta_{\rm eff}$ trivially coincides with $\beta$. In particular, in Ref.\,\cite{GherardiniPRE2018}, the energy exchanged between a two-level quantum system and a measurement apparatus was analyzed, with the assumption that the repeated interaction with the measurement device can be reliably modeled by a sequence of projective measurements occurring instantly and at random times. Numerically, it has been observed that, as compared with the case of measurements occurring at fixed times, the two-level system exchanges more energy in the presence of randomness when the average time between consecutive measurements is sufficiently small in comparison with the inverse resonance frequency. However, the quantum-heat Jarzynski equality, related to the equilibrium properties of the transformation applied to the system, is still obeyed, as well as when the waiting times between consecutive measurements are randomly distributed and for each random realization. These results are theoretically supported by the fact that, in the analyzed case, the dynamical evolution of the quantum system is unital\,\cite{Rastegin13JSTAT13,Sagawa2014}. A discussion on the values of the parameter $\beta_{\rm eff}$, extracted from experimental data for nitrogen-vacancy (NV) centers in diamonds subject to projective measurements in a regime where an effective two-level approximation is valid was recently presented in Ref.\,\cite{Hernandez2019}.

\begin{figure}[h!]
    \centering
    \includegraphics[scale=0.5]{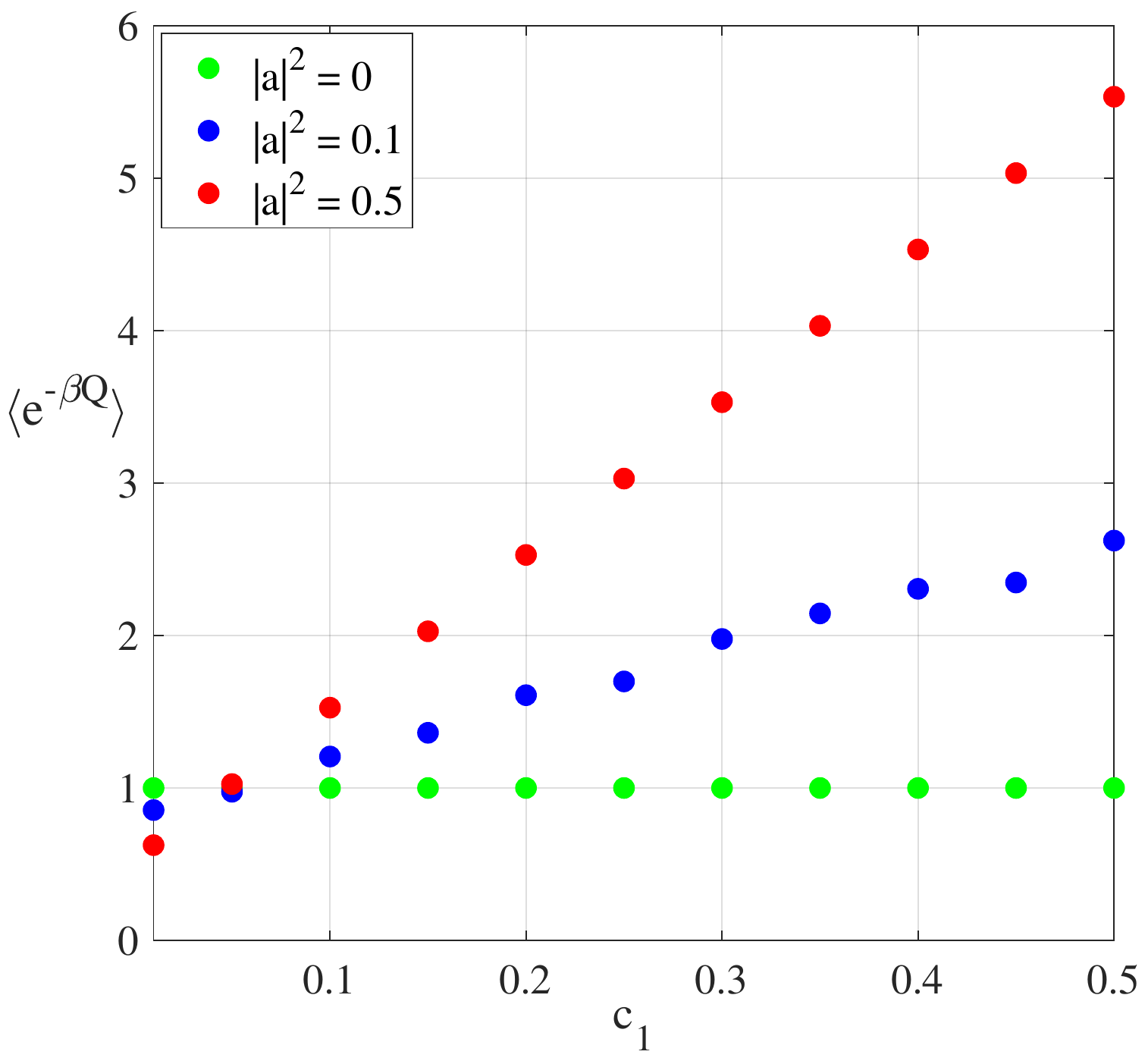}
    \caption{Quantum-heat characteristic function $\langle e^{-\beta Q}\rangle$ for a two-level quantum system as a function of $c_1$ in Equation (\ref{c1}) for three values of $|a|^2$, which characterizes the initial state. The function is obtained from numerical simulations performed for a system with a Hamiltonian $H= J(\ketbra{0}{1} + \ketbra{1}{0})$ subject to a sequence of $M=5$ projective measurements. The latter are separated by a fixed waiting time $\tau=0.5$, averaged over $2000$ realizations, with $E_{1,2}=\pm 1$ and $\beta=3/2$ (notice that the same values of $|a|^2$ are used in Fig.\,1 of \cite{GherardiniPRE2018} and in the corresponding caption the value of $\beta$ should read $\beta=3/2$). Units are used with $\hbar=1$ and $J=1$.}
    \label{fig:G_2LS}
\end{figure}

In Figure \ref{fig:G_2LS}, we plot the quantum-heat characteristic function $\langle e^{-\beta Q}\rangle$ as a function of the parameter $c_1$ that appears in the decomposition of the initial state $\rho_0$ with respect to the energy eigenstates $|E_1\rangle$ and $|E_2\rangle$

\begin{equation}\label{c1}
\rho_0 = c_1\ketbra{E_1}{E_1} + c_2\ketbra{E_2}{E_2} \ ,
\end{equation}
where $c_2 = 1 - c_1$. The function is plotted for three values of the parameter $|a|^2$, used to parametrize the eigenstates $\{|\Omega_1\rangle,|\Omega_2\rangle\}$ of $O$ as a function of the energy eigenstates of the system, i.e.,
\begin{equation}
|\Omega_1\rangle = a\ket{E_1} - b\ket{E_2}\,\,\,\,\,\text{and}\,\,\,\,\,
|\Omega_2\rangle = b\ket{E_1} + a\ket{E_2} \ ,
\end{equation}
with $|a|^2 + |b|^2 = 1$ and $a^{\ast}b = ab^{\ast}$. As a result, one can observe that $\langle e^{-\beta Q}\rangle = 1$ for the value of $c_1$ ensuring that $\rho_0=e^{-\beta H}/Z$. Further details can be found in Ref.\,\cite{GherardiniPRE2018}.

The $N>2$ case is trickier: Since, in general, the initial state is no longer thermal, it is not trivial to determine the value of $\beta_{\rm eff}$, and its dependence on the initial condition is interesting to investigate. In this regard, in the following, we will numerically address the $N=3$ case by providing results on the asymptotic behavior of the system in the limit $M\gg 1$. For the sake of simplicity, from here on, we assume that the values of the waiting times $\tau_i$ are fixed for any $i=1,\ldots,M$. However, our findings turn out to be the same for any choice of the marginal probability distribution functions $p(\tau_i)$, up to ``pathological'' cases such as $p(\tau) = \delta(\tau)$. So, having random waiting times does not significantly alter the scenario emerging from the presented results.

\section{Parametrization of the Initial State}\label{sec:param}

In this paragraph, we introduce a parametrization of the initial state $\rho_0$ for the $N=3$ case, which will be useful for a thermodynamic analysis of the system.

As previously recounted, for a two-level quantum system in a mixed state, as given by Equation \eqref{inital}, it is always formally possible to define a temperature. In particular, one can implicitly find an effective inverse temperature, making $\rho_0$ a thermal state, by solving the following equation for $\beta \in \mathbb{R}$
\begin{equation} \label{beta2livelli}
e^{-\beta(E_2-E_1)} = \frac{c_2}{c_1} \ .
\end{equation}

In the $N=3$ case, however, we have two independent parameters in \eqref{inital} (the third parameter indeed enforces the condition $\text{Tr}[\rho_0] = 1$), and thus, in general, it is no longer possible to formally define a single temperature for the state. Here, we propose the following parametrization of $c_1, c_2, c_3$ that generalizes the one of Equation \eqref{beta2livelli}.

We denote as \emph{partial} effective temperatures the three parameters $b_1,b_2,b_3$, defined through the ratios of $c_1$, $c_2$, and $c_3$
\begin{equation}
\frac{c_2}{c_1} = e^{- b_1 (E_2-E_1)},\hspace{1cm}\frac{c_3}{c_2} = e^{- b_2 (E_3-E_2)}, \hspace{1cm} \frac{c_1}{c_3} = e^{- b_3 (E_1-E_3)} \ ,
\end{equation}
such that, for a thermal state, $b_k = \beta$, $\forall k=1,2,3$. The three parameters are not independent, as they are constrained by the relation
\begin{equation}
\frac{c_2}{c_1} \frac{c_3}{c_2} \frac{c_1}{c_3} = 1 \ ,
\end{equation}
which gives in turn the following equality
\begin{equation}\label{eq:constraint}
b_1 (E_2-E_1) + b_2 (E_3-E_2) + b_3 (E_1-E_3) = 0 \ .
\end{equation}

By introducing $\Delta_1 = E_2 - E_1$, $\Delta_2 = E_3 - E_2$, and $\Delta_3 = E_1 - E_3$, Equation \eqref{eq:constraint} can be written as
\begin{equation}\label{orthogonality}
\sum^3_{k=1} b_k \Delta_k = 0 \ ,
\end{equation}
where by definition
\begin{equation}\label{plane}
\sum^3_{k=1} \Delta_k = 0 \ .
\end{equation}

Thus, as expected, the thermal state is a solution of the condition \eqref{orthogonality} for any choice of $E_1,E_2,E_3$. This has also a geometric interpretation. In the space of the $\Delta_k$, $k=1,2,3$, Equation \eqref{orthogonality} becomes an orthogonality condition between the $\Delta_k$ and the $b_k$ vectors, while \eqref{plane} defines a plane that is orthogonal to the vector $(1,1,1)$. When $b_k$ is proportional to $(1,1,1)$, the orthogonality condition is automatically satisfied and one finds a thermal state. This suggests that, in general, one can conveniently parametrize $b_k$ in terms of both the components that are orthogonal and parallel to the plane $\sum^3_{k=1} \Delta_k = 0 $. Such terms have the physical meaning of the thermal and non-thermal components of the initial state. Formally, this means that we can parametrize each $b_k$ through the fictitious inverse temperature $\beta$ and a deviation $\alpha$, i.e.,
\begin{equation}
(b_1, b_2, b_3) = \beta (1,1,1) + \frac{\alpha}{v} (\Delta_3 - \Delta_2,\,\Delta_1 - \Delta_3,\,\Delta_2 - \Delta_1)  \ ,
\end{equation}
where $v$ acts as a normalization constant
\begin{equation}
v^2 = 3\left
(\Delta_1^2 + \Delta_2^2 + \Delta_3^2\right) \ .
\end{equation}

Hence, taking into account the normalization constraint, the coefficients $c_k$ are given by
\begin{equation}
c_1 = \frac{1}{1 + e^{-b_1 \Delta_1} + e^{b_3 \Delta_3}}, \hspace{1cm} c_2 = \frac{1}{1 + e^{-b_2 \Delta_2} + e^{b_1 \Delta_1}}, \hspace{1cm} c_3 = \frac{1}{1 + e^{-b_3 \Delta_3} + e^{b_2 \Delta_2}},
\end{equation}
or, in terms of the parameters $\alpha$ and $\beta$,
\begin{equation}\label{eq:coeffs_c}
c_1 = \frac{1}{\Tilde{Z}} \exp\left[-\beta E_1 + \frac{\alpha}{v} (E_2 - E_3)^2\right], \hspace{0.4cm} c_2 = \frac{1}{\Tilde{Z}} \exp\left[-\beta E_2 + \frac{\alpha}{v}(E_3 - E_1)^2\right], \hspace{0.4cm} c_3 =\frac{1}{\Tilde{Z}} \exp\left[- \beta E_3 + \frac{\alpha}{v}(E_1 - E_2)^2\right] ,
\end{equation}
where $\Tilde{Z}$ is a pseudo-partition function ensuring the normalization of the $c_k$'s
\begin{equation} \label{pseudoZ}
\Tilde{Z} = \Tilde{Z} (\alpha, \beta) \equiv e^{-\beta E_1 + \frac{\alpha}{v} (E_2 - E_3)^2} +  e^{- \beta E_2 + \frac{\alpha}{v}(E_3 - E_1)^2} + e^{- \beta E_3 + \frac{\alpha}{v}(E_1 - E_2)^2} \ .
\end{equation}

Let us provide some physical intuition about the parameters $\alpha$ and $\beta$: For $\alpha = 0$, we recover a thermal state, whereby $c_1 > c_2 > c_3$ if $\beta > 0$, or vice versa if $\beta < 0$. On the other hand, the non-thermal component $\alpha$ can be used to obtain a non-monotonic behavior of the coefficients $c_k$. For example, for $\beta=0$, since $(E_3 -E_1)^2$ is greater than both $(E_3 - E_2)^2$ and $(E_1 - E_2)^2$, one finds that $c_2 >(<)\,c_1,c_3$ if $\alpha >(<)\,0$.

As a final remark, it is worth noting that we can reduce the dimension of the space of the parameters. In particular, without loss of generality, one can choose the zero of the energy by taking $E_2=0$ (and then $E_3>0$, $E_1 <0$), or we can reduce our analysis to the cases with $\beta > 0$. As a matter of fact, the parametrization is left unchanged by the transformation $\{\beta \rightarrow - \beta,\,E_k \rightarrow - E_k\}$, with the result that the case of $\beta < 0$ can be explored by simply considering $\beta >0$ in the fictitious system with $E^{\prime}_k = - E_k$ (here, the choice of $E_2=0$ guarantees that this second case can be simply obtained by substituting $E_1$ with $E_3$).

\section{Large $M$ Limit}

Here, we numerically investigate the behavior of a three-level system subject to a sequence of $M$ projective quantum measurements with a large $M$ (asymptotic limit) and where $\tau$ is not infinitesimal. From here on, we adopt the language of spin-$1$ systems, and we thus identify $O$ with $S_z$.

In the asymptotic limit, the behavior of the system is expected not to depend on the choice of the evolution Hamiltonian, with the exception that at least one of the eigenstates of $S_z$ is also an energy eigenstate. In such a case, indeed, if the energy outcome corresponding to the common eigenstate is obtained by the first measurement of the TPM scheme, then the evolution is trivially deterministic, as the system is locked in the measured eigenstate.

\begin{figure}[h!]
    \centering
    \includegraphics[scale=0.56]{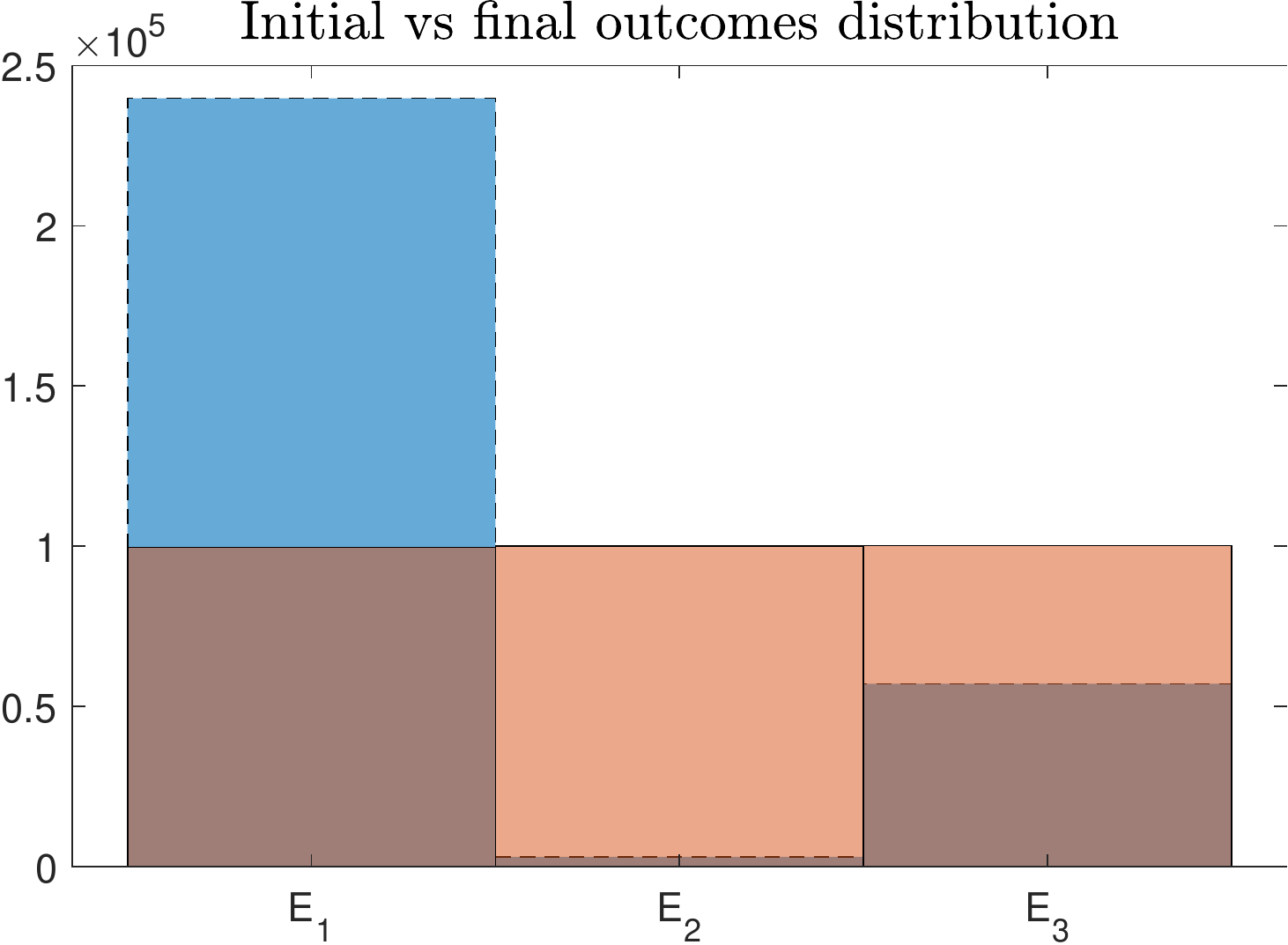}
    \caption{Histogram of the initial (dashed) and final (solid) energy outcomes for the TPM scheme described in the text performed over $3 \cdot 10^5$ realizations. While the initial state is non-uniform, the final state is practically uniform over the three energy levels.}
    \label{fig:istogramma}
\end{figure}

Choosing a generic observable (with no eigenstates in common with $H$), numerical simulations (cf.\,Figure \ref{fig:istogramma}) suggest that our protocol leads the system to the completely uniform state. The latter can be interpreted as a canonical state with $\beta = 0$ (notice that this result holds in the situation we are analyzing, with
a finite dimensional Hilbert space). The system evolves with Hamiltonian $H= \omega_1 S_z + \omega_2 S_x$, where the energy units are chosen such that $\omega_1=1$ and $\omega_2 = \frac{1}{2}$. It is initialized in the state $\rho_0$ with $\{c_1 = 0.8,c_2 = 0.01,c_3 = 0.19\}$, corresponding to $\alpha \approx - 2,32 $ and $\beta \approx 1,96$, and we performed $M=20$ projective measurements of the observable $O=S_z$ separated by the time $\tau=1$.

This numerical finding allows us to derive an analytic expression of the quantum-heat characteristic function. In this regard, as the final state is independent of the initial one for a large $M$, the joint probability of obtaining $E_n$ and $E_m$, after the first and the second energy measurement respectively, is equal to
\begin{equation}
p_{mn} = \frac{1}{3} c_m \ .
\end{equation}

Hence,
\begin{equation} \label{G}
G(\epsilon) = \me{e^{- \epsilon Q}} = \frac{1}{3} \sum^3_{m,n=1} c_m e^{- \epsilon(E_m-E_n)} = \frac{1}{3} \sum_{n=1}^3 e^{- \epsilon E_n} \sum_{m=1}^3 c_m e^{ \epsilon E_m} \ .
\end{equation}

As a consequence, $G$ can be expressed in terms of the partition function $Z(\beta)$ and of the pseudo-partition function introduced in Equation $\eqref{pseudoZ}$, i.e.
\begin{equation} \label{GZ}
G(\epsilon;\alpha,\beta) = \frac{Z(\epsilon)}{Z(0)} \frac{\Tilde{Z}(\alpha, \beta - \epsilon)}{\Tilde{Z}(\alpha,\beta)} \ .
\end{equation}

\begin{figure}[h!]
    \centering
    \subfloat[][]{\includegraphics[scale=0.54]{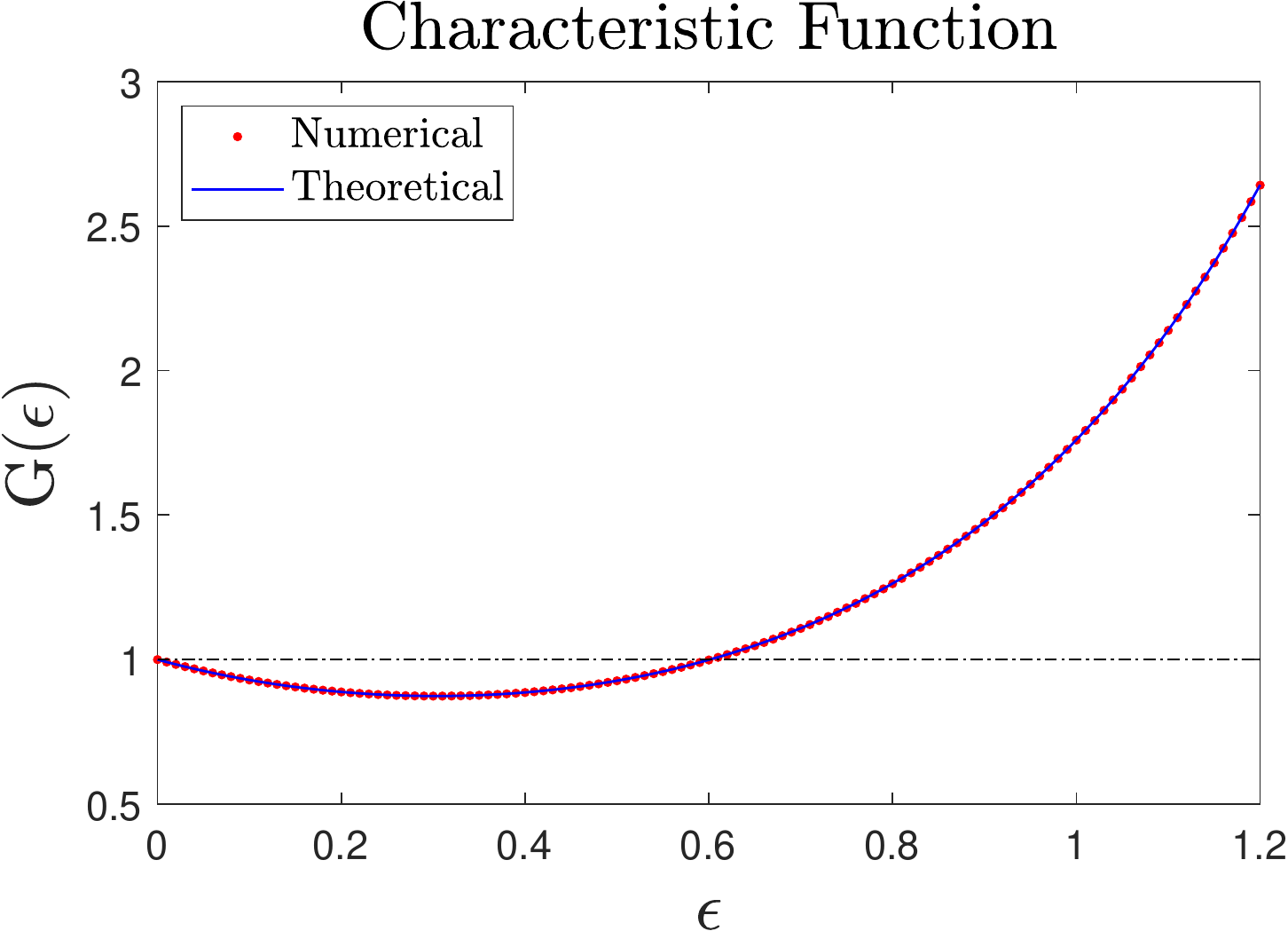}}
    \subfloat[][]{\includegraphics[scale=0.54]{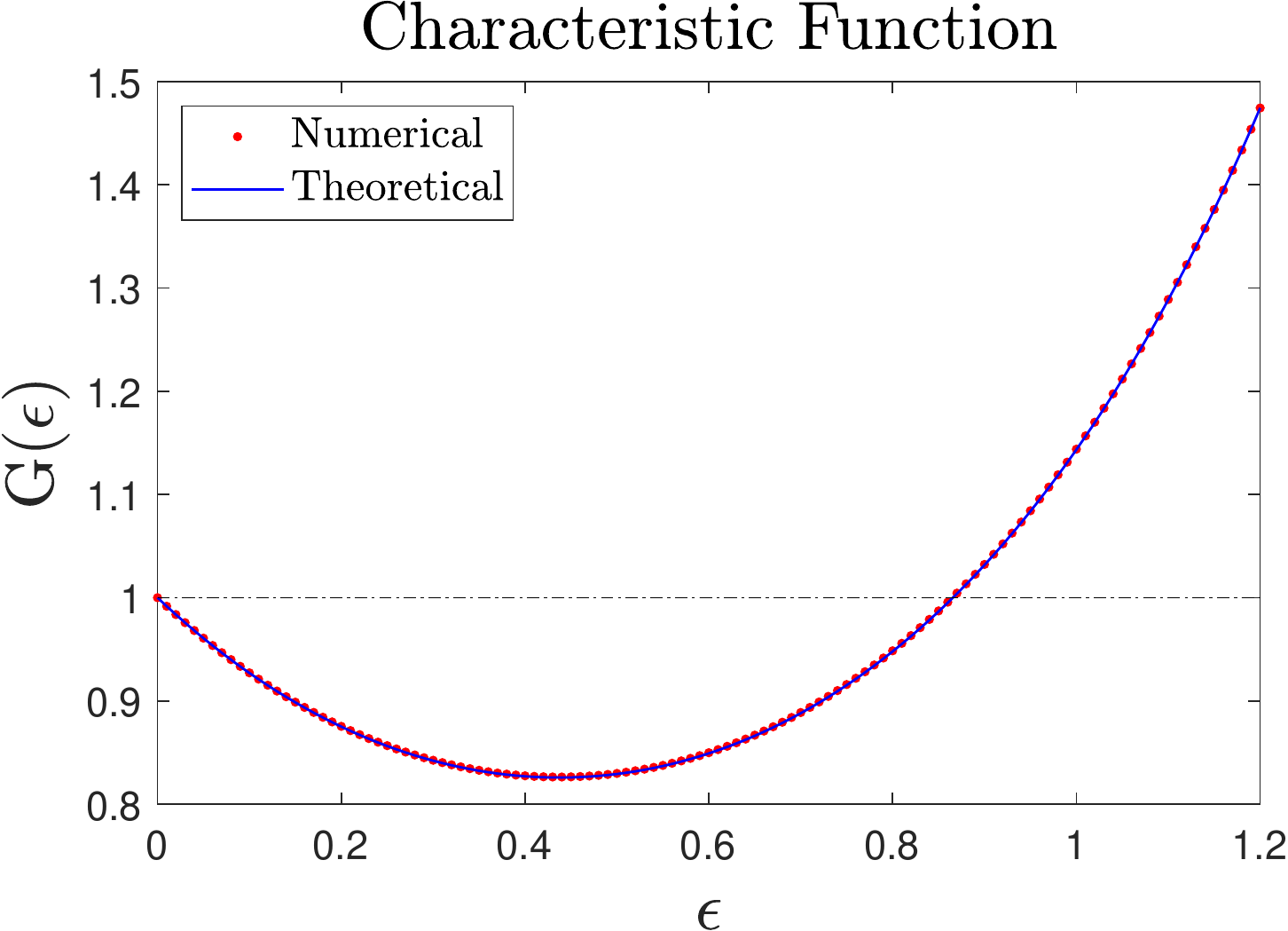}}
    \caption{{Comparison of the analytic expression} \eqref{G} of the asymptotic (large-$M$) quantum-heat characteristic function $G(\epsilon)$ (blue solid lines) with the numerical results averaged over $3 \cdot 10^5$ realizations (red dots). The initial state is the same as in Figure \ref{fig:istogramma} and, again, $O=S_z$. In panel (\textbf{a}), the Hamiltonian is the same as in Figure  \ref{fig:istogramma},
    while in panel (\textbf{b}) the Hamiltonian is $H= \omega_1 S^2_z + \omega_2 S_x$, with $\omega_1=2\omega_2=1$.}
    \label{fig:G}
\end{figure}

Regardless of the choice of the system parameters, the already known results are straightforwardly recovered, i.e., $G(0)=1$ and $G(\beta)=1$ for $\alpha=0$ (initial thermal state). In Figure \ref{fig:G}, our analytical expression for $G(\varepsilon)$ is compared with its numerical estimate for two different Hamiltonians, showing a very good agreement.

We remark that the distribution of $\rho$ after the second energy measurement could also be obtained by simply imposing the maximization of the von Neumann entropy. This is reasonable, since the measurement device is macroscopic and can provide any amount of energy. As a final remark, notice that, in the numerical findings of Figure \ref{fig:G}, the statistics of the quantum-heat fluctuations originated by the system respect the same ergodic hypothesis that is satisfied whenever a sequence of quantum measurements is performed on a quantum system\,\cite{Gherardini2016NJP,Gherardini2017QSc,PiacentiniNatPhys2017}. In particular, in Figure \ref{fig:G}, one can observe that the analytical expression of $G(\epsilon)$ for a large $M$ (i.e., in the asymptotic regime obtained by indefinitely increasing the time duration of the implemented protocol) practically coincides with the numerical results obtained by simulating a sequence with a finite number of measurements ($M=20$) but over a quite large number ($3 \cdot 10^5$) of realizations. This evidence is quite important, because it means that the quantum-heat statistics is homogeneous and fully take into account even phenomena occurring with very small probability in a single realization of the protocol.

\section{Estimates of $\beta_{\rm eff}$}

\begin{figure}[h!]
    \centering
    \subfloat[]{\includegraphics[scale=0.6]{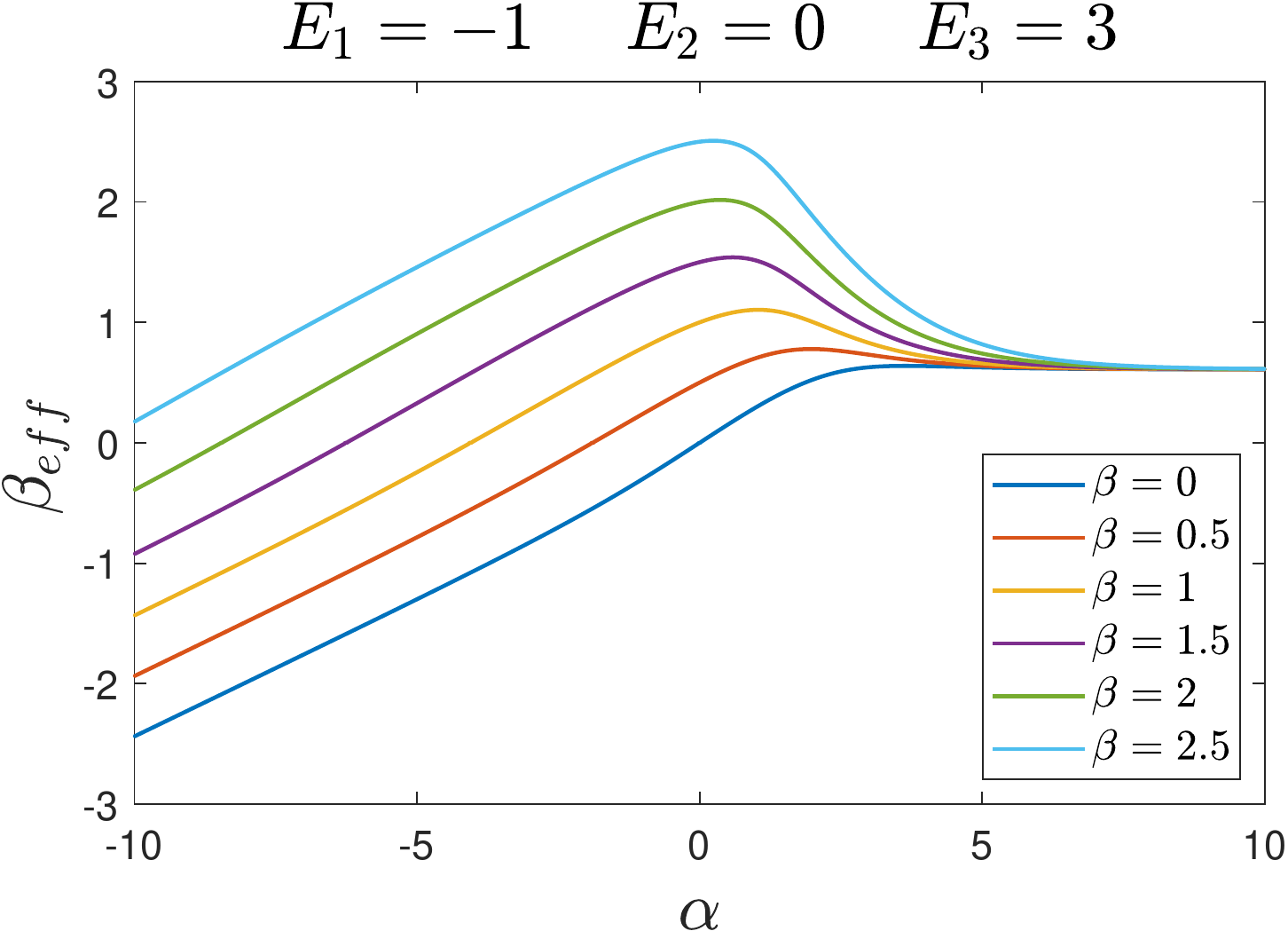}} \\
    .\ \\
    \subfloat[]{\includegraphics[scale=0.612]{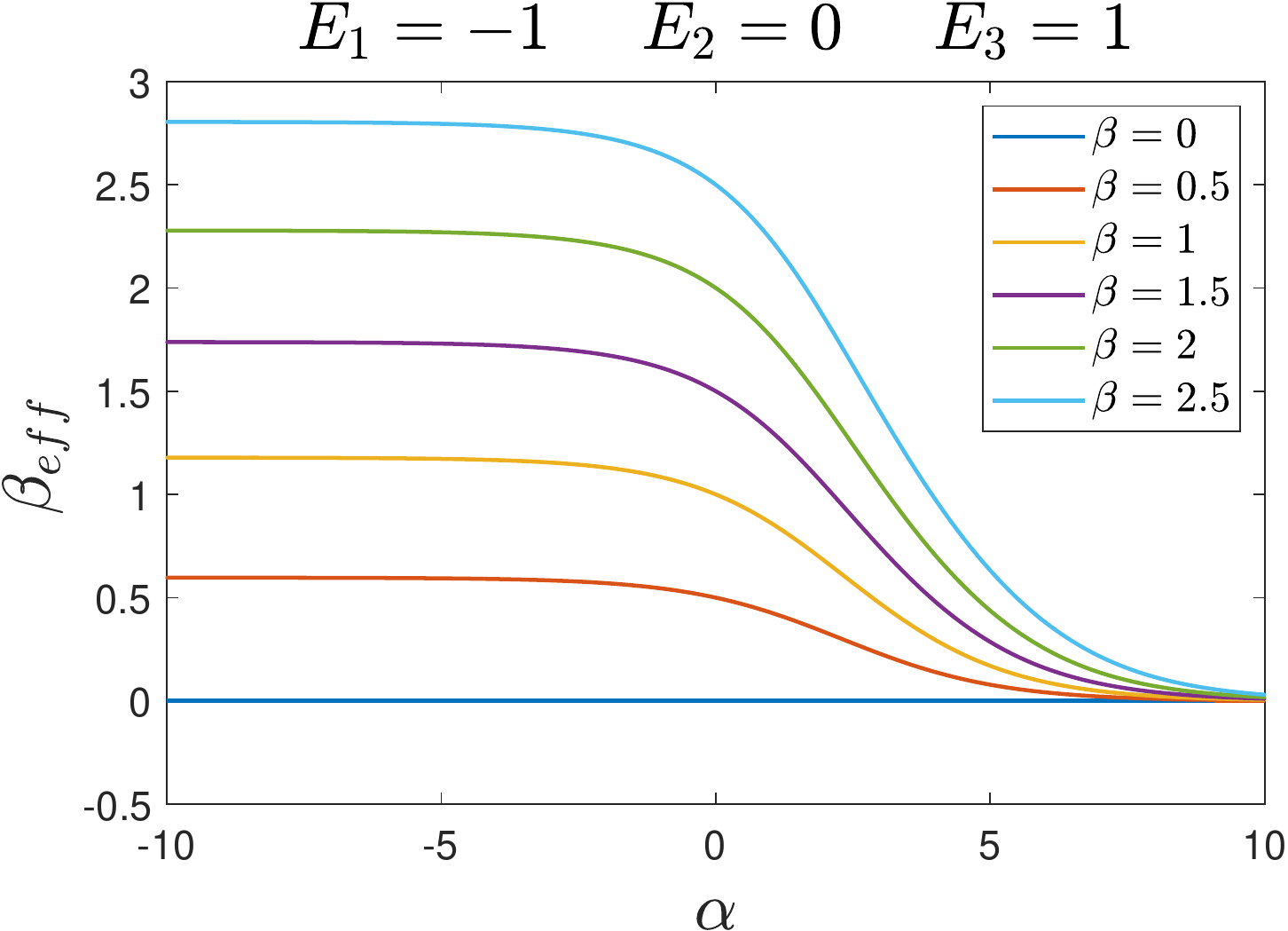}} \\
    .\ \\
   \subfloat[]{\includegraphics[scale=0.6]{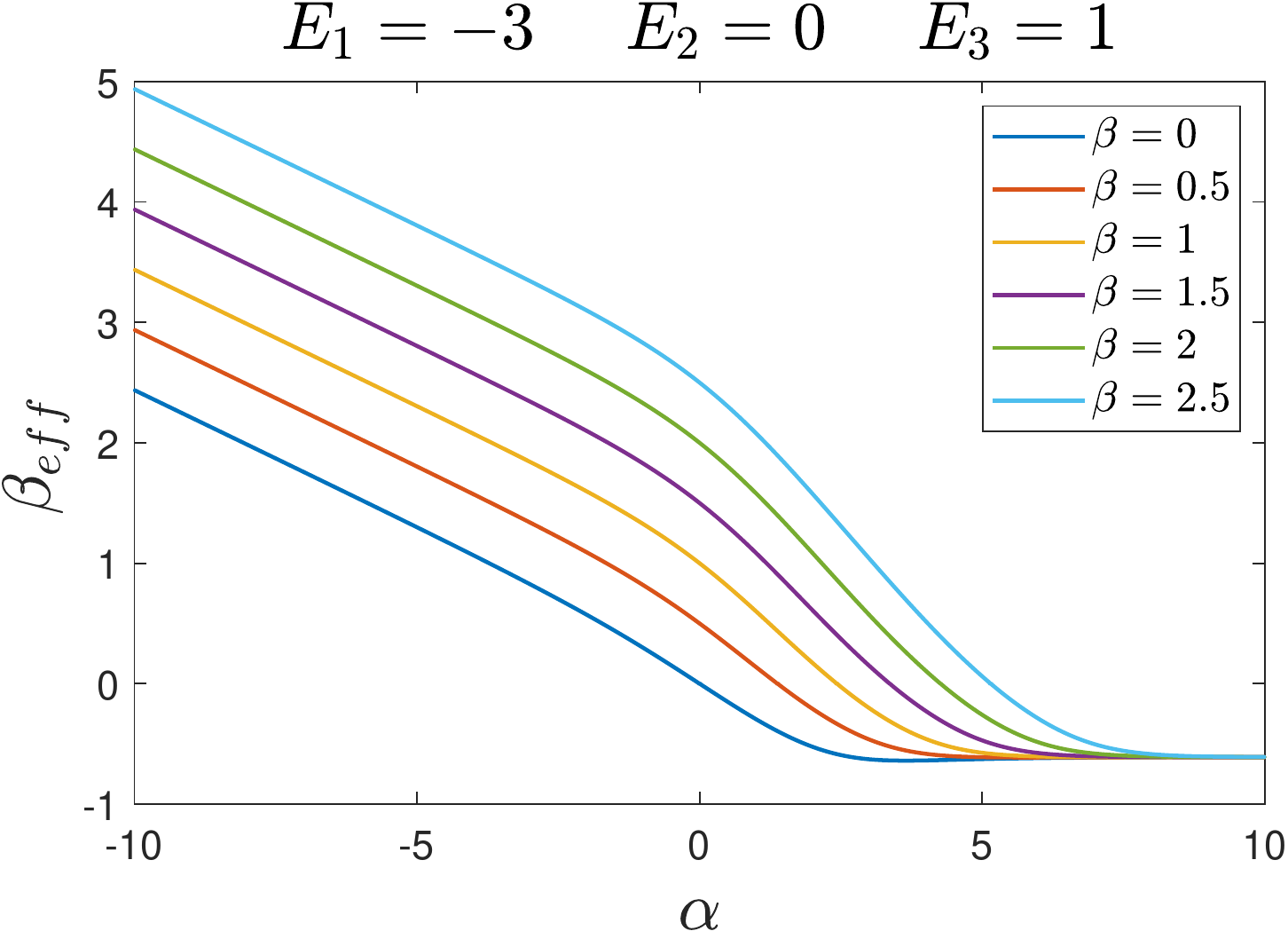}}
    \caption{Behavior of $\beta_{\rm eff}$ as a function of $\alpha$ for different values of $\beta \in [0,2.5]$. We have chosen: (\textbf{a}) $\{E_1 = -1,\,E_2 = 0,\,E_3 = 3\}$, (\textbf{b}) $\{E_1 = -1,\,E_2 = 0,\,E_3 = 1\}$, and (\textbf{c}) $\{E_1 = -3,\,E_2 = 0,\,E_3 = 1\}$, respectively.}
    \label{fig:betaeff}
\end{figure}

In this section, we study the behavior of $\beta_{\rm eff}$, i.e., the nontrivial solution of $G(\beta_{\rm eff}) = 1 $, as a function of the initial state (parametrized by $\alpha$ and $\beta$) and of the energy levels of the system. Let us first notice that, by starting from Equation \eqref{G},
obtaining an analytical expression for $\beta_{\rm eff}$ in the general case appears to be a very non-trivial task.
Thus, in Figure \ref{fig:betaeff}, we numerically compute $\beta_{\rm eff}$ as a function of $\alpha$ (the non-thermal component of $\rho_0$) for different values of $\beta$. Three representative cases for the energy levels are taken, i.e., $\{E_1 = -1,\,E_2 = 0,\,E_3 = 3\}$, $\{E_1 = -1,\,E_2 = 0,\,E_3 = 1\}$, and $\{E_1 = -3,\,E_2 = 0,\,E_3 = 1\}$, respectively. This choice allows us to deal both with the cases $E_3 - E_2 > E_2 - E_1$ and $E_3 - E_2 < E_2 - E_1$. The choice of the energy unit is such that the smallest energy gap between $E_3 - E_2$ and $E_2 - E_1$ is set to one. As stated above, we consider $\beta >0$; the corresponding negative values of the inverse temperature are obtained by taking $E^{\prime}_k = - E_k$ with $\beta^{\prime} = - \beta$. As expected, for $\alpha = 0$, we have $\beta_{\rm eff} = \beta$, regardless of the values of the $E_k$'s.

In the next two subsections, we continue discussing in detail the findings of Figure \ref{fig:betaeff},
presenting the asymptotic behaviors for large positive and negative values
of $\alpha$.

\subsection{Asymptotic Behavior for a Large Positive $\alpha$}

From Figure \ref{fig:betaeff}, one can deduce that, for large positive values of $\alpha$ (corresponding to having as the initial density operator the pure state $\rho_0 = \ketbra{E_2}{E_2}$), $\beta_{\rm eff} \rightarrow \bar{\beta}_{\rm eff}$, which only depends on $E_1$ and $E_3$. This asymptotic value $\bar{\beta}_{\rm eff}$ is positive if $E_3 - E_2 > E_2 - E_1$, negative if $E_3 - E_2 < E_2 - E_1$, and zero when $E_3 - E_2 = E_2 - E_1$. To better explain the plots in Figure \ref{fig:betaeff}, let us consider the analytic expression of $G(\epsilon)$. In this regard, for a large $\alpha$ and finite $\beta$, we can write
\begin{equation}
\Tilde{Z}(\alpha, \beta) \approx  e^{ - \beta E_2 + \frac{\alpha}{v}(E_3 - E_1)^2} \ ,
\end{equation}
so that, by using Equation \eqref{GZ}, the condition $G(\beta_{\rm eff}) = 1$ reads as $e^{-\bar{\beta}_{\rm eff} (E_1 - E_2)} + e^{-\bar{\beta}_{\rm eff}(E_3-E_2)} = 2$, or, setting $E_2=0$,
\begin{equation} \label{asymptotic}
e^{-\bar{\beta}_{\rm eff} E_1} + e^{ -\bar{\beta}_{\rm eff} E_3} = 2 \ .
\end{equation}

Notice that, if $E_3 = - E_1$ the only solution of Equation \eqref{asymptotic} is $\bar{\beta}_{\rm eff} =0$, while a positive solution appears for $E_3 > - E_1$ and a negative one if $E_3 < - E_1$, thus confirming what was observed in the numerical simulations. Moreover, by replacing $E_1 \rightarrow - E_3$ and $E_3 \rightarrow - E_3$, the value of $\bar{\beta}_{\rm eff}$ changes its sign.

Now, without loss of generality, let us fix the energy unit so that $E_1 =-1$. The behavior of $\bar{\beta}_{\rm eff}$ as a function of $E_3$ is shown in Figure \ref{fig:asymptotic}. We observe a monotonically increasing behavior of $\bar{\beta}_{\rm eff}$ up to a constant value for $E_3 \gg |E_1| =1$. Once again, this value can be analytically computed from Equation \eqref{asymptotic}, which, for a large value of $E_3$, gives $\bar{\beta}_{\rm eff} = \ln{2}$. Putting together all of the above considerations and restoring the energy scales, the limits of $\bar{\beta}_{\rm eff}$ are the following
\begin{equation} \label{limits}
- \frac{\ln{2}}{E_3 - E_2} < \bar{\beta}_{\rm eff} < \frac{\ln{2}}{E_2-E_1} \ .
\end{equation}

The lower and the upper bounds of $\bar{\beta}_{\rm eff}$ are also shown in Figure \ref{fig:asymptotic}, in which $E_1=-1$ and $E_2 = 0$.

\subsection{Asymptotic Behavior for a Large Negative $\alpha$}

From Figure \ref{fig:betaeff}, one can also conclude that, for large negative values of $\alpha$, the behavior of $\beta_{\rm eff}$ is linear with $\alpha$
\begin{equation} \label{linear}
\beta_{\rm eff} \approx r \alpha \ ,
\end{equation}
with $r>0$ if $E_3 - E_2 > E_2 - E_1$, $r=0$ when $E_3 - E_2 = E_2 - E_1$, and $r$ is negative otherwise. This divergence is easily understood: In fact, the limit $\alpha \rightarrow - \infty$ (for finite $\beta$) corresponds to the initial state $\rho_0 = \ketbra{E_1}{E_1}$ when $E_3 - E_2 < E_2 - E_1$ or $\rho_0 = \ketbra{E_3}{E_3}$ if $E_3 - E_2 > E_2 - E_1$. On the other hand, those states (thermal states with $\beta_{\rm eff} = \beta = \pm \infty$) are also reached in the limits $\beta \rightarrow \pm \infty$ with a finite $\alpha$. This simple argument does not imply the linear divergence of $\beta_{\rm eff}$ as in Equation \eqref{linear}, nor does it provide insights
about the value of $r$, which, however, can be derived from Equation \eqref{GZ}.
Although the calculation makes a distinction on the sign of $r$ depending on whether $E_3 - E_2$ is greater or smaller than $E_2 - E_1$, the result is independent of this detail. In particular, by considering the case $E_3 - E_2 > E_2 - E_1$ ($r>0$) and taking into account the divergence of $\beta_{\rm eff} = r \alpha$, we find in the $\alpha \rightarrow - \infty$ regime that the characteristic function $G(\beta_{\rm eff})$ has the following form:
\begin{equation}
G(\beta_{\rm eff}) = \frac{1}{3} + \text{const} \times e^{- \alpha |\Delta_3|\left[r-\frac{(\Delta_1 - \Delta_2)}{v}\right]} \ .
\end{equation}
\begin{figure} [h|]
    \centering
    \subfloat[]{\includegraphics[scale=0.55]{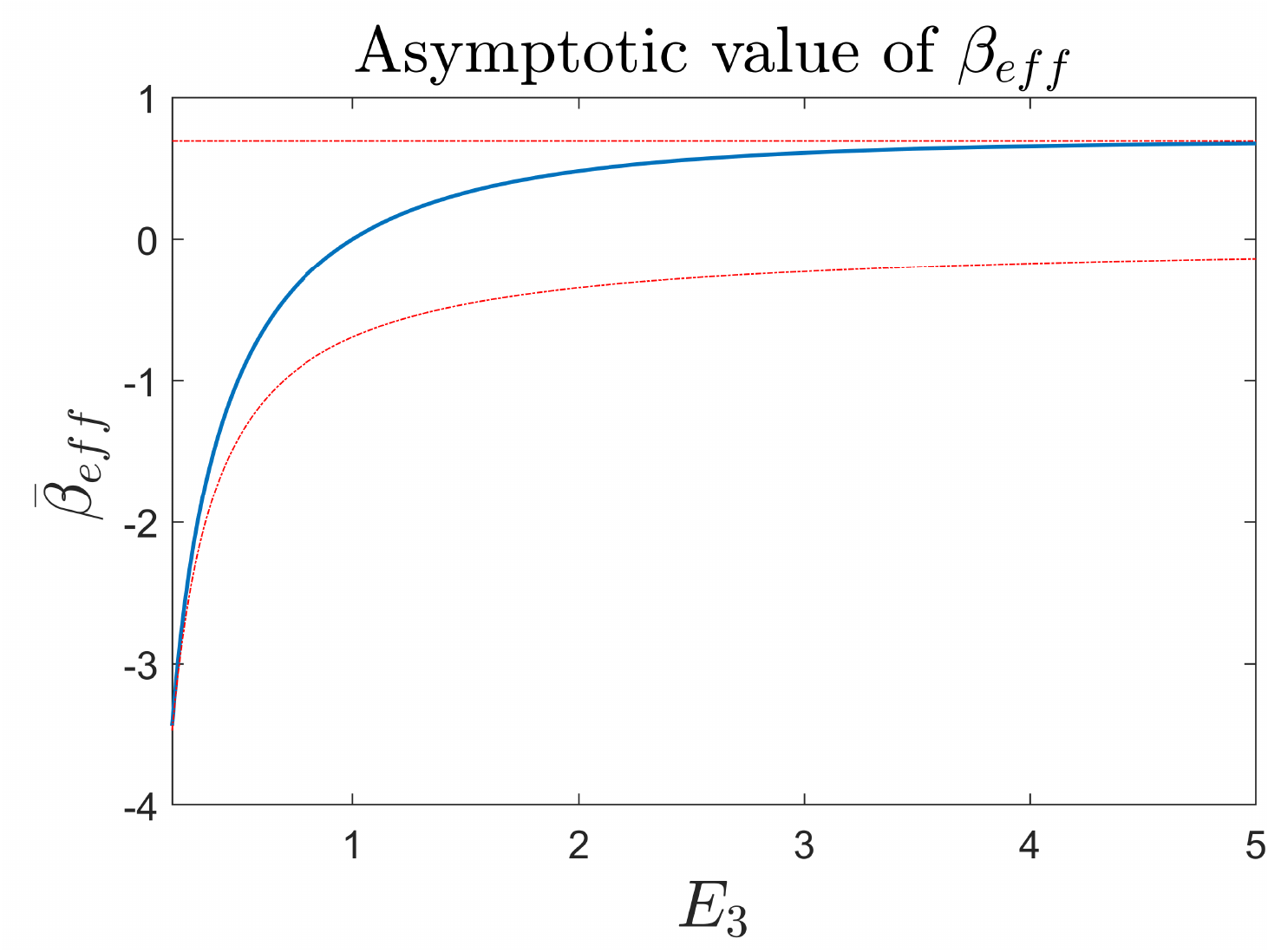}} \ \
    \subfloat[]{\includegraphics[scale=0.55]{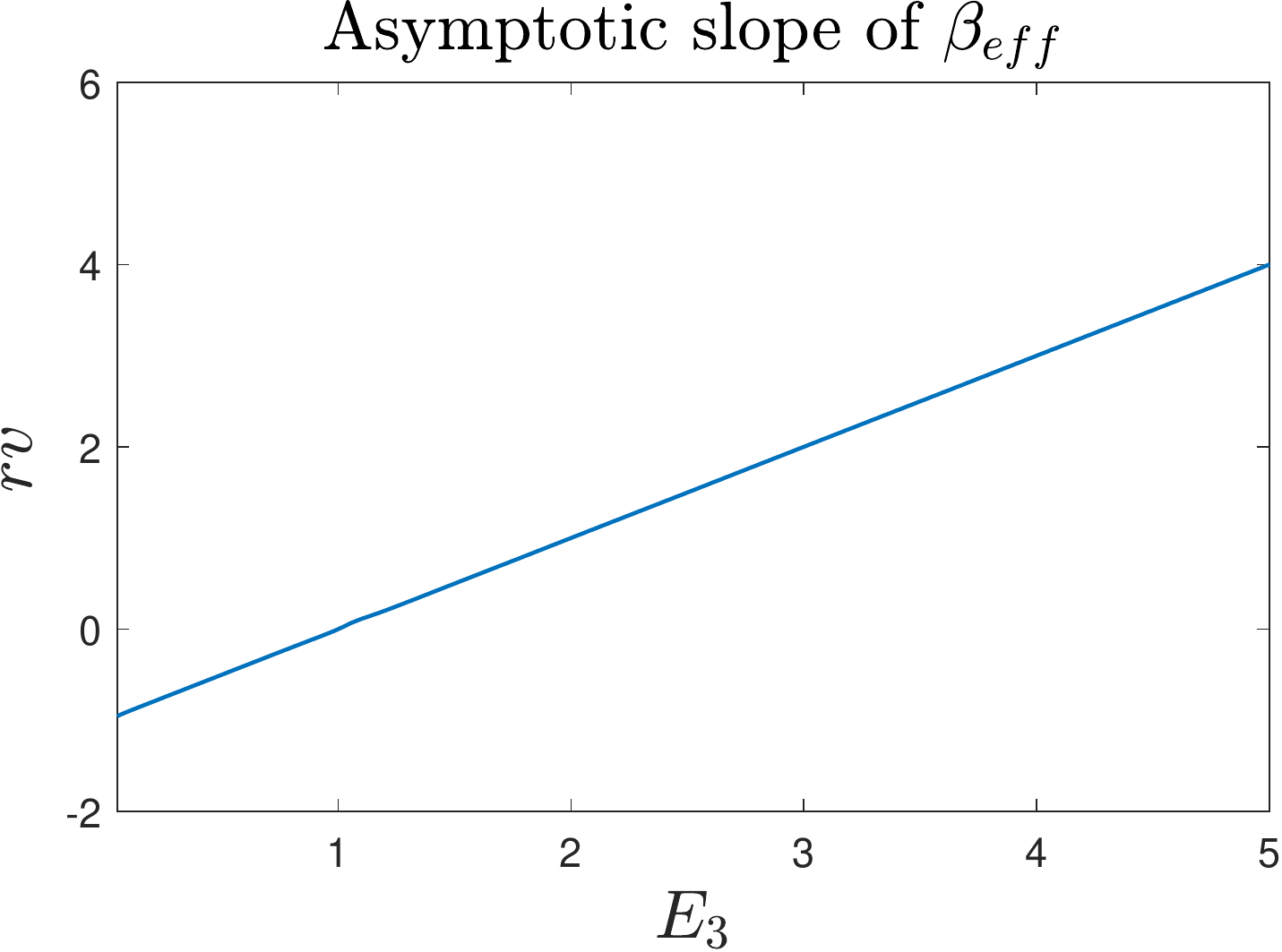}}
    \caption{(\textbf{a}) Behavior of the asymptotic value $\bar{\beta}_{\rm eff}$ for a large positive $\alpha$ ($\alpha = 20$) as a function of $E_3$ (solid blue line) with $E_1 = -1$ and $E_2 = 0$. We compare the curve with its limiting (lower and upper) values, defined in Equation \eqref{limits} (dash-dotted red lines). (\textbf{b}) Behavior of the asymptotic slope $r$, rescaled for $v$, for a large negative $\alpha$ ($\alpha = -20$) as a function of $E_3$. In both cases,
    $E_2=0$ and $E_1 = -1$.}
    \label{fig:asymptotic}
\end{figure}

Hence, in order to ensure that $G(\beta_{\rm eff}) \neq \frac{1}{3}$ in the limit $\alpha \rightarrow - \infty$, the following relation has to be satisfied
\begin{equation}
r = \frac{E_1 + E_3 - 2 E_2}{v} \ .
\end{equation}

The numerical estimate of $rv$ as a function of $E_3$ is shown in Figure \ref{fig:asymptotic}. The numerical results confirm the linear dependence of $\beta_{\rm eff}$ as a function of $\alpha$.

\subsection{Limits of the Adopted Parametrization}

\begin{figure}[h!]
    \centering
    \subfloat[]{\includegraphics[scale=0.55]{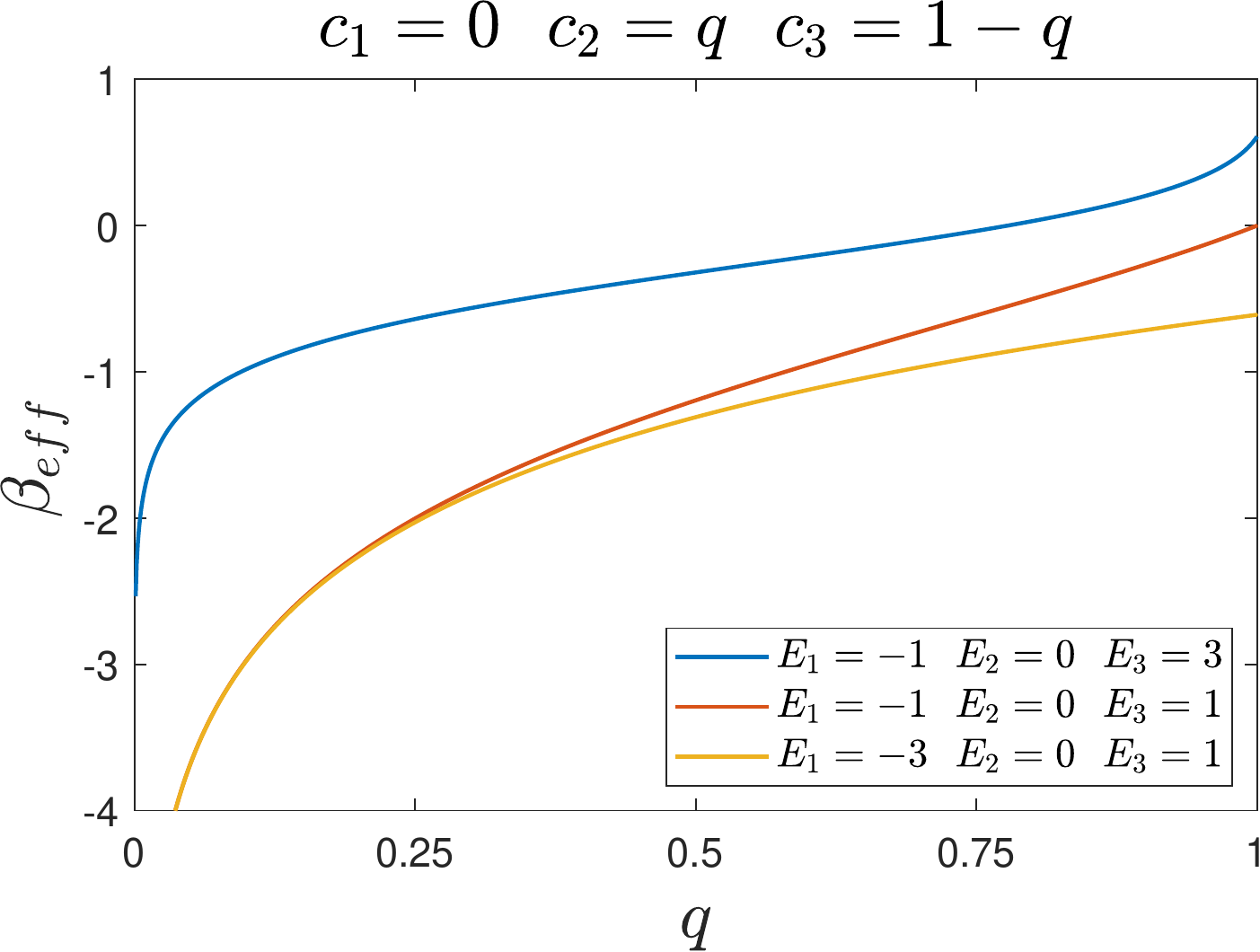}} \\
    \subfloat[]{\includegraphics[scale=0.55]{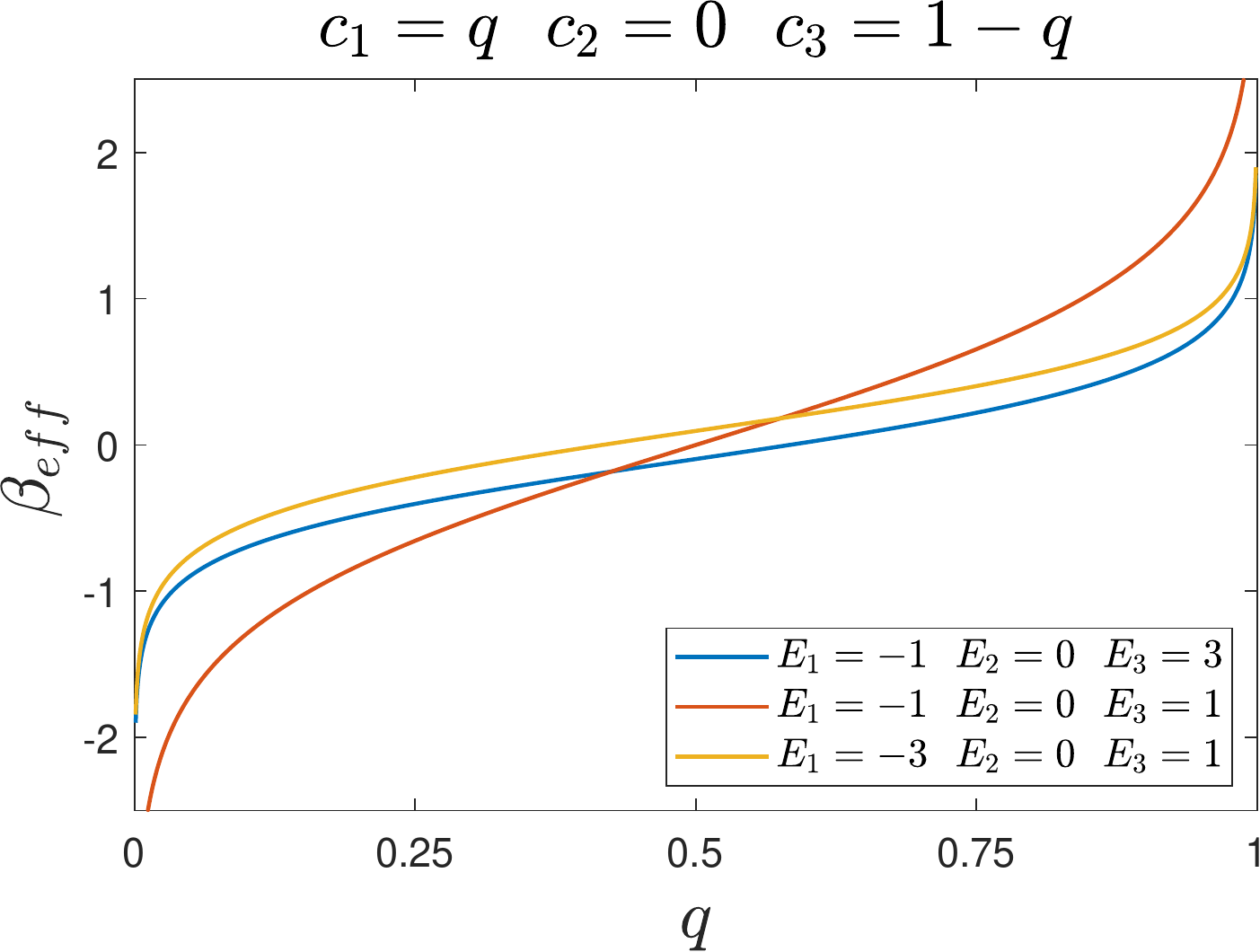}} \\
   \subfloat[]{\includegraphics[scale=0.55]{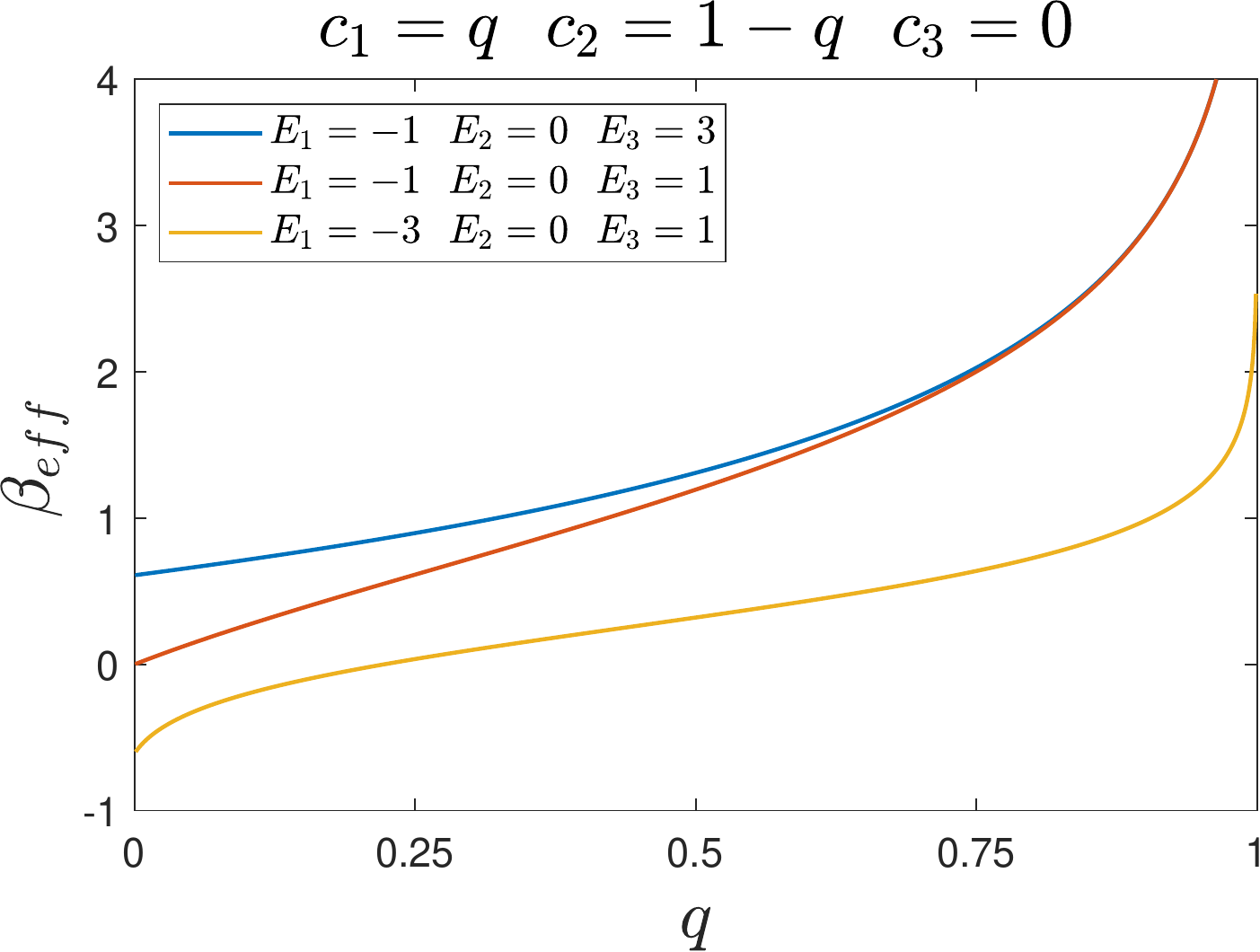}}
    \caption{{Behavior of} $\beta_{\rm eff}$ as a function of $q$, which parametrizes the initial state $\rho_0$ as in Equation \eqref{c=0}, in each of the three cases (\textbf{a}) $E_3 - E_2 > E_2 - E_1$, (\textbf{b}) $E_3 - E_2 = E_2 - E_1$, and (\textbf{c}) $E_3 - E_2 < E_2 - E_1$.}
    \label{fig:betaeff_2}
\end{figure}

The parametrization introduced in Section \ref{sec:param} is singular in correspondence of the
initial states $\rho_0$ with one or more coefficients $c_k$ equal to zero. In this regard, as remarked above, initial pure states can be easily obtained in the limits $\beta \rightarrow \pm \infty$, $\alpha$ finite (corresponding to $\rho_0 = \ketbra{E_1}{E_1}$ and $\rho_0 = \ketbra{E_3}{E_3}$, respectively) and $\alpha \rightarrow + \infty$, $\beta$ finite (that provides $\rho_0 = \ketbra{E_2}{E_2}$). Instead, initial states with only a coefficient $c_k$ equal to zero, namely
\begin{equation}\label{c=0}
\rho_0 = q \ketbra{E_1}{E_1} + (1-q) \ketbra{E_2}{E_2} \ , \hspace{0.5 cm} \rho_0 = q \ketbra{E_1}{E_1} + (1-q) \ketbra{E_3}{E_3} \ , \hspace{0.5cm} \rho_0 = q \ketbra{E_2}{E_2} + (1-q) \ketbra{E_3}{E_3} \ ,
\end{equation}
with $q \in [0,1]$, cannot be easily written in terms of $\alpha$ and $\beta$. In fact, the expressions in Equation (\ref{c=0}) correspond to the limit in which $\alpha \rightarrow - \infty$ with $\beta = a \alpha + b$ for suitable $a$, $b$. This result can be obtained, e.g., for the first of the states in Equation \eqref{c=0}, considering the state $c_1 = q (1-e^{-Y})$, $c_2 = e^{-Y}$, and $c_3 = (1-q)(1-e^{-Y})$ in the limit $Y \rightarrow + \infty$. Solving for $\alpha$ and $\beta$, we have
\begin{equation}
\alpha = - \frac{v}{\Delta_1 \Delta_2} Y + O(1) \ , \hspace{1cm}    \beta = - \frac{r}{3} \frac{v}{\Delta_1 \Delta_2} Y + O(1) \ ,
\end{equation}
so that $a = r/3$, while the $q$ dependence is encoded in the next-to-leading term. For this reason, the parametrization in terms of $q \in [0,1]$ turns out to be the most convenient in the case of singularity. In Figure \ref{fig:betaeff_2}, the numerical estimates of $\beta_{\rm eff}$ as a function of $q$ are shown for the three cases in Equation \eqref{c=0}, respectively for $E_3 - E_2$ greater, equal to, and smaller than $E_2 - E_1$. The symmetries $E_1 \rightarrow - E_3$, $E_3 \rightarrow - E_1$, $q \rightarrow 1- q$ and $\beta_{ \rm eff} \rightarrow -\beta_{\rm eff}$, due to our choice of  parametrization, can be observed.

\section{Conclusions}

In this paper, we studied the quantum-heat distribution originating from a three-level quantum system subject to a sequence of projective quantum measurements.

As figure of merit, we analyze the characteristic function $G(\epsilon)=\langle e^{-\epsilon Q}\rangle$ of the quantum heat $Q$ by using the formalism of stochastic thermodynamics. In this regard, it is worth recalling that, as the system Hamiltonian $H$ is time-independent, the fluctuations of the energy variation during the protocol can be effectively referred of as quantum heat.
As shown in Ref.\,\cite{Hernandez2019}, the fluctuation relation describing all the statistical moments of $Q$ is simply given by the equality $G(\beta_{\rm eff})=1$, where the energy-scaling parameter $\beta_{\rm eff}$ can be considered as an effective inverse temperature. The analytic expression of $\beta_{\rm eff}$ has been determined only for specific cases, as, for example, two-level quantum systems.

Here, a three-level quantum system was considered and, in order to gain information on the value of $\beta_{\rm eff}$, we performed specific numerical simulations. In doing this, we introduced a convenient parametrization 
of the initial state $\rho_0$, such that its population values can be expressed as a function of the reference inverse temperature $\beta$ and the parameter $\alpha$, identifying the deviation of $\rho_0$ from the thermal state. Then, the behavior of the system when $M$, the number of projective measurements, is large is numerically analyzed. The condition of a large $M$ leads to an asymptotic regime whereby the final state of the system tends to a completely uniform state, stationary with respect to the energy basis. This means that such a state can be equivalently described by an effective thermal state with zero inverse temperature. In this regime, the value of the energy scaling $\epsilon$ allowing for the equality $G(\epsilon)=1$ (i.e., $\beta_{\rm eff}$) is evaluated. As a consequence, $\beta_{\rm eff}$, which uniquely rescales energy exchange fluctuations, implicitly encloses information on the initial state $\rho_0$. In other terms, once $\rho_0$ and the system (time-independent) Hamiltonian $H$ are fixed, $\beta_{\rm eff}$ remains unchanged by varying parameters pertaining to the measurements performed during the dynamics, e.g., the time interval between the measurements.

We have also determined $\beta_{\rm eff}$ as a function of $\alpha$ and $\beta$ for large $M$. Except for few singular cases, we found that, for large negative values of $\alpha$, $\beta_{\rm eff}$ is linear with respect to $\alpha$, while it tends to become constant and independent of $\beta$ for large positive values of $\alpha$. Such conditions can be traced back to an asymptotic equilibrium regime, because any dependence from the initial state $\rho_0$ is lost.

As a final remark, we note that, overall, the dynamics acting on the analyzed three-level system are unital\,\cite{Rastegin13JSTAT13,Sagawa2014}. As a matter of fact, this is the result of a non-trivial composition of unitary evolutions (between each couple of measurements) and projections. It would certainly also be interesting to analyze a three-level system subject to both a sequence of quantum measurements and in interaction with an external (classical or quantum) environment. In this respect, in light of the results in Refs.\,\cite{WoltersPRA2013,Hernandez2019}, the most promising platforms for this kind of experiment are NV centers in diamonds\,\cite{DohertyPhysRep2013}. Finally, also the analysis of general $N$-level systems, and the study of large $M$ behavior deserve further investigations.

\subsection*{Acknowlegments}

The authors gratefully acknowledge M. Campisi, P. Cappellaro, F. Caruso, F. Cataliotti, N. Fabbri, S. Hern\'andez-G\'omez, M. M\"uller and F. Poggiali for useful discussions. This work was financially supported by the MISTI Global Seed Funds MIT-FVG Collaboration Grant ``NV centers for the test of the Quantum Jarzynski Equality (NVQJE)''. The author (SG) also acknowledges PATHOS EU H2020 FET-OPERN grant No. 828946 and UNIFI grant Q-CODYCES. The author (SR) thanks the editors of this issue for inviting him to write a paper in honour of Shmuel Fishman, whom he had the pleasure to meet several times and appreciate his broad and deep knowledge of various fields of the theory of condensed matter; besides that, Shmuel Fishman was a lovely person with whom it was a privilege to spend time in scientific and more general discussions.

\end{document}